\documentclass[amsmath, amssymb, showpacs, showkeys, pra]{revtex4}
\usepackage[english]{babel}
\usepackage{graphicx}
\usepackage{bm}

\def\pr{^{\,\prime}}
\def\ds{\displaystyle}

\def\erfc{\,\mathrm{erfc}}
\def\erf{\,\mathrm{erf}}
\def\ei{\,\mathrm{Ei}}

\def\const{\,\mathrm{const}}

\def\A{\,\mathcal{A}}

\def\S{\,\mathsf{S}}
\def\SI{\,\mathcal{S}}
\def\F{\,\mathsf{F}}
\def\I{\,\mathcal{I}}

\def\sh{\,\mathrm{sinh}}
\def\ch{\,\mathrm{cosh}}
\def\th{\,\mathrm{tanh}}
\def\laplace{\Delta}

\begin{document}

\title{Model for Diffusion-Induced Ramsey Narrowing}

\author{Alexander Romanenko}
\email{alexrm@univ.kiev.ua}
\affiliation{Physics faculty, Kyiv National Taras Shevchenko University, pr.Glushkova 6, 03680, Kyiv, Ukraine}

\author{Leonid Yatsenko}
\email{yatsenko@iop.kiev.ua}
\affiliation{Institute of Physics NAS of Ukraine, pr.Nauki 46, 03028, Kyiv, Ukraine\\}

\date{\today}

\begin{abstract}
Diffusion-induced Ramsey narrowing that appears when atoms can leave the interaction region and repeatedly return without lost of coherence is investigated using strong collisions approximation. The effective diffusion equation is obtained and solved for low-dimensional model configurations and three-dimensional real one.
\end{abstract}


\maketitle

\section{Introduction} 
\label{introduction}
\begin{figure}
\includegraphics[scale=0.4]{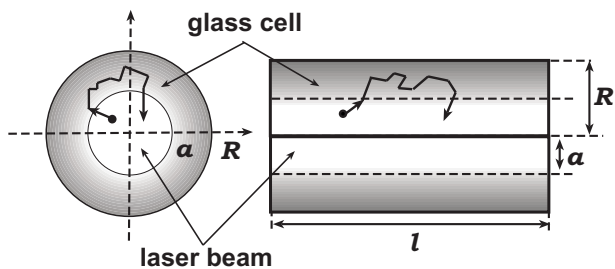}
\caption{Scheme of the experiment}
\label{fig:scheme}
\end{figure}

The effect of diffusion-induced Ramsey narrowing has been considered in~\cite{xiao}.
It appears in the system of atoms moving through a laser beam in a buffer gas.
The lifetime of the atomic coherence is limited and leads to the Lorentz line shape 
in assumption that atoms diffuse out of the laser beam and do not return.
However, after spending some time in dark (outside the laser beam),
atoms can return to the interaction region before losing the coherence (see fig.~(\ref{fig:scheme})).
In some cases when decoherence effects are small diffusing atoms can spend 
a majority of their coherence time in the dark which leads to spectral narrowing of the center 
of the atomic resonance line shape in analogy to Ramsey spectroscopy~\cite{ramsey}.

In the present paper we propose the model for diffusion-induced Ramsey narrowing
based on kinetic equation in approximation of strong collisions. The paper is organized as follows.
In section~\ref{sect1} we discuss the basic kinetic equation in strong collision approximation.
Further we obtain the effective diffusion equation in section~\ref{sect2} and solve it exactly 
in section~\ref{sect3} for the simplest case unbounded configuration.
Using initial approximation we obtain the similar diffusion equation for 
spatially bounded configuration in~\ref{sect4}. Short review of results 
is given in section~\ref{sect5}.

\section{Basic Equations} 
\label{sect1}
In the classical problem of migration of a particle in the gas
the kinetic equation for the distribution function~$\rho(\vec{t}, \vec{v}, t)$
has the form (\cite{sobelman},~\cite{rautian}):
\begin{equation}
\frac{\partial\rho}{\partial t}+(\vec{v}\cdot\vec{\nabla})\rho=\SI\,,
\label{basic:boltzmann}
\end{equation}
where $\SI$ is the collision integral that can be expressed in terms of~$\rho$
as follows:
\begin{equation}
\SI=-\int
\bigl[ 
\A(\vec{v}, \vec{v}')\rho(\vec{r}, \vec{v}, t)
-\A(\vec{v}', \vec{v})\rho(\vec{r}, \vec{v}', t)
\bigr]\,d\vec{v}'\,.
\label{basic:collision1}
\end{equation}
Here $\A(\vec{v}, \vec{v}')$ means the numbers of transition $\vec{v}\to\vec{v}'$ 
during the unit of time due to collisions with the buffer gas.
In this expression collisions between atoms are neglected. The initial condition is 
$\rho(\vec{r}, \vec{v}, 0)=W(\vec{v})\delta(\vec{r})$, where $W(\vec{v})$ is Maxwell distribution.

The collision integral can be written as
\begin{equation}
\SI=
-\nu\rho(\vec{r}, \vec{v}, t)+
\int
\bigl[ 
-\A(\vec{v}', \vec{v})\rho(\vec{r}, \vec{v}', t)
\bigr]\,d\vec{v}'\,,
\label{basic:collision2}
\end{equation}
where 
\begin{equation}
\nu(\vec{v})=\int\A(\vec{v}, \vec{v}')\,d\vec{v}'
\label{basic:collision3}
\end{equation}
denote the effective collision frequency.

There are two simple models that allow one to investigate the kinetic equation~(\ref{basic:boltzmann})
with the collision integral ~(\ref{basic:collision2}). The first one based on the approximation of weak collisions
(heavy atoms scatter on light particles) and leads to diffusion approximation with Fokker-Planck equation for the function~$\rho$.
In this model the velocity of the atom changes essentially after great number of collisions.

The second model based on the approximation of strong collisions 
(light atoms scatter on heavy particles). In this model the function $\A(\vec{v}', \vec{v})$ is independent of $\vec{v}'$, 
this means that the particle velocity after collision $\vec{v}$ is independent 
of its velocity before the collision.
The form of $\A(\vec{v})$ can be determined from the condition $\SI=0$ in statical equilibrium.
In this case $\nu=\const$, and 
\begin{equation*}
\A(\vec{v})=\nu W(\vec{v})\,.
\end{equation*}
This expression means that an arbitrary initial distribution turn to Maxwell 
after the lapse of time~$1/\nu$ (or, equivalently, the atom forgets its initial velocity 
as time~$1/\nu$ passes).

Finally, the kinetic equation for the problem of migration of the atom in gas 
in approximation of strong collisions has the form:
\begin{equation*}
\frac{\partial \rho}{\partial t}+(\vec{v}\cdot\vec{\nabla})\rho
=-\nu\rho+\nu W(\vec{v})\int\rho(\vec{r}, \vec{v}, t)\,d\vec{v}\,.
\end{equation*}

The same approach can be used (see~\cite{rautian}) for non-diagonal elements of density matrix.
For two-level system we have:
\begin{equation*}
\frac{\partial \rho_{12}}{\partial t}+(\vec{v}\cdot\vec{\nabla})\rho_{12}
=-(\nu+\gamma+i\Delta\omega)\rho_{12}+\nu W(\vec{v})\int\rho_{12}(\vec{r}, \vec{v}, t)\,d\vec{v}+q(\vec{r}, \vec{v})\,,
\end{equation*}
where $\gamma$ describes a spontaneous decay, 
$q(\vec{r}, \vec{v})$ is coherent excitation (with laser beam), $\Delta\omega$ is two-photon frequency detuning.

We will consider the stationary case. 
Denoting $\rho(\vec{r}, \vec{v})=\rho_{12}(\vec{r}, \vec{v})$
we can write:
\begin{equation}
(\vec{v}\cdot\vec{\nabla})\rho(\vec{r}, \vec{v})=-(\nu+\gamma+i\Delta\omega)\rho(\vec{r}, \vec{v})
+W(\vec{v})
\left[ 
\lambda(\vec{r})+
\nu\int d\vec{v}\,\rho(\vec{r}, \vec{v})
\right]\,,
\label{basic:equation}
\end{equation}
where
\begin{equation*}
W(\vec{v})=W_{0}e^{-\vec{v}\,^{2}/v_{0}^{2}}\,,\quad
\lambda(\vec{r})=\lambda_{0}e^{-\vec{r}\,^{2}/a^{2}}\,.
\end{equation*} 
and $W_{0}$ is the normalization constant for the Maxwell distribution.

Let us denote
\begin{equation*}
N(\vec{r})=\int d\vec{v}\,\rho(\vec{r}, \vec{v})\,,\quad
\alpha=\nu+\gamma+i\Delta\omega\,,\quad
\alpha_{0}=\gamma+i\Delta\omega\,.
\label{basic:signal}
\end{equation*}
The complex signal is defined as:
\begin{equation}
\S(\Delta\omega)=\iint d\vec{r}\,d\vec{v}\,\lambda(\vec{r})\rho(\vec{r}, \vec{v})
=\int N(\vec{r})\lambda(\vec{r})\,d\vec{r}\,.
\end{equation}
We assume that
\begin{equation*}
\Delta\omega, \gamma\ll\nu\,,\quad
\nu\gg\frac{v_{0}}{a}=\frac{1}{\tau_{0}}\,,
\end{equation*}
where $\tau_{a}=\frac{a}{v_{0}}$~--- 
is characteristic radiation zone time of flight.
The last relations can be rewritten as
$\tau_{a}\gg \tau_{\nu}=\frac{1}{\nu}$ (collision time).

For the atomic ensemble the atomic resonance line shape is determined by weighted average 
of the line shapes from different Ramsey sequences (with alternating interactions with the laser beam and motion in the dark).


\section{Effective Diffusion Equation}
\label{sect2}
According to~(\ref{basic:signal}) the complex signal~$\S(\Delta\omega)$ can be expressed in terms
of zeroth moment $N(\vec{r})$ of the distribution $\rho(\vec{r})$ with respect to $\vec{v}$. 
Let us obtain the equations for higher moments $N^{(k)}(\vec{r})$. For the sake of simplicity 
first we consider 1-dimensional case, where
\begin{equation*}
N^{(k)}(x)=\int\limits_{-\infty}^{\infty}v^{k}\rho(x, v)\,dv\,,\quad N^{(0)}(x)=N(x)\,.
\end{equation*}
Multiplying the equation 
\begin{equation*}
v\frac{\partial }{\partial x}\rho(x, v)=-\alpha\rho(x, v)+W(v)\bigl[ \lambda(x)+\nu N^{(0)}(x)\bigr]
\end{equation*}
by $v^{k}$ and integrating over $v$ we get
\begin{equation*}
\frac{dN^{(k+1)}}{dx}=-\alpha N^{(k)}(x)+\langle v^{k}\rangle \bigl[ \lambda(x)+\nu N^{(0)}(x)\bigr]\,,\quad
\langle v^{k}\rangle=\int\limits_{-\infty}^{\infty}v^{k}W(v)\,dv\,.
\end{equation*}
Since $\langle v^{k}\rangle=0$ for odd values of $k$, we can write the following equations for odd and even orders:
\begin{equation*}
\frac{dN^{(2k+1)}}{dx}=-\alpha N^{(2k)}+\langle v^{2k}\rangle \bigl[ \lambda(x)+\nu N^{(0)}(x)\bigr]\,,\\
\quad
\frac{dN^{(2k+2)}}{dx}=-\alpha N^{(2k+1)}
\end{equation*}
or, equivalently equations for even orders only:
\begin{equation}
\frac{1}{\alpha}\frac{d^{2}N^{(2k+2)}}{dx^{2}}=\alpha N^{(2k)}-\langle v^{2k}\rangle \bigl[ \lambda(x)+\nu N^{(0)}(x)\bigr]\,,\\
\label{basic:moments}
\end{equation}
In these sequence of equations one can substitute the next equation to the previous and obtain the expression like the following:
\begin{equation}
\alpha N^{(0)}-\bigl[ \lambda(x)+\nu N^{(0)}(x)\bigr]
=\frac{1}{\alpha}\frac{d^{2}N^{(2)}}{dx^{2}}
=\frac{\langle v^{2}\rangle}{\alpha^{2}}\frac{d^{2}}{dx^{2}}\bigl[ \lambda(x)+\nu N^{(0)}(x)\bigr]+\frac{1}{\alpha^{3}}\frac{d^{4}N^{(4)}}{dx^{4}}=\dots
\label{basic:expansion1}
\end{equation}
Performing the substitution $k$ times we obtain:
\begin{equation*}
\alpha N^{(0)}=
\left( 
1+\frac{\langle v^{2}\rangle}{\alpha^{2}}\frac{d^{2}}{dx^{2}}
+\dots
+\frac{\langle v^{2k}\rangle}{\alpha^{2k}}\frac{d^{2k}}{dx^{2k}}
\right)\bigl[ \lambda(x)+\nu N^{(0)}(x)\bigr]+
\frac{1}{\alpha^{2k+1}}\frac{d^{2k+2}N^{(2k+2)}}{dx^{2k+2}}\,.
\end{equation*}
For $k\to\infty$ this expression can be written symbolically:
\begin{equation}
\alpha N^{(0)}=\sum\limits_{k=0}^{\infty}\frac{\langle v^{2k}\rangle}{\alpha^{2k}}\frac{d^{2k}}{dx^{2k}}
\bigl[ \lambda(x)+\nu N^{(0)}(x)\bigr]=\hat{\F}\bigl[ \lambda(x)+\nu N^{(0)}(x)\bigr]\,,
\label{basic:operator:form}
\end{equation}
where $\hat{\F}$ is a formal linear operator defined as
\begin{equation*}
\hat{\F}=\int\limits_{-\infty}^{\infty}dv\,W(v)\biggl( 1-\frac{v}{\alpha}\frac{d}{dx}\biggr)^{-1}
\end{equation*}

For higher dimensions the result is similar but more complicated because the moments became tensors:
\begin{equation*}
N^{(k)}_{i_{1}\dots i_{k}}(\vec{r})=\int v_{i_{1}}\dots v_{i_{k}} \rho(\vec{r}, \vec{v})\,d\vec{v}
\end{equation*}
In 3-dimensional case the same procedure leads to
\begin{equation*}
\nabla_{j}N^{(2k)}_{i_{1}\dots i_{2k-1}j}=-\alpha N^{(2k-1)}_{i_{1}\dots i_{2k-1}}\,,\quad
\nabla_{j}N^{(2k+1)}_{i_{1}\dots i_{2k}j}=-\alpha N^{(2k)}_{i_{1}\dots i_{2k}}
+\langle v_{i_{1}}\dots v_{i_{2k}}\rangle (\lambda+\nu N^{(0)})\,.
\end{equation*}
Hence, 
\begin{equation*}
N^{(0)}=\left( 
1+\frac{1}{\alpha^{2}}\langle v_{i}v_{j}\rangle \nabla_{i}\nabla_{j}
+\frac{1}{\alpha^{2}}\langle v_{i}v_{j}v_{k}v_{l}\rangle \nabla_{i}\nabla_{j}\nabla_{k}\nabla_{l}+\dots
\right)(\lambda+\nu N^{(0)})\,.
\end{equation*}
Since
\begin{equation*}
\langle v_{i}v_{j}\rangle=\frac{1}{3}\,\langle v^{2}\rangle\delta_{ij}\,,\quad
\langle v_{i}v_{j}v_{k}v_{l}\rangle
=\frac{1}{15}\,\langle v^{4}\rangle
\bigl( \delta_{ij}\delta_{kl}+\delta_{ik}\delta_{jl}+\delta_{il}\delta_{jk}\bigr)\,,
\end{equation*}
we can write finally
\begin{equation}
N^{(0)}=\left( 
1+\frac{\langle v^{2}\rangle}{3\alpha^{2}}\laplace+
\frac{\langle v^{4}\rangle}{5\alpha^{4}}
\laplace^{2}+\dots\right)(\lambda+\nu N^{(0)})\,.
\label{basic:expansion2}
\end{equation}
The expressions~(\ref{basic:expansion1}) and~(\ref{basic:expansion2}) are asymptotic expansions 
and for sufficiently large $\nu$ (and $\alpha$) we can leave only first terms. The terms with second-order derivatives 
lead to the diffusion equation. It is easily to find that these terms have the same coefficients for 
all dimensions:
\begin{equation*}
\alpha N=(\lambda+\nu N)+\frac{v_{0}^{2}}{2\alpha^{2}}\laplace(\lambda\nu N)+\dots
\end{equation*}
or one can write up to the terms $1/\nu$:
\begin{equation}
\alpha_{0} N(\vec{r})=
\frac{\nu v_{0}^{2}}{2\alpha^{2}}\,\Delta N(\vec{r})+\lambda(\vec{r})
\label{diffusion:equation}
\end{equation}
This equation can be treated as diffusion equation with absorption factor~$\alpha_{0}=\gamma+i\Delta\omega$ and 
diffusion coefficient 
\begin{equation}
D=\frac{\nu v_{0}^{2}}{2\alpha^{2}}
\sim\frac{v_{0}^{2}}{2\nu}
\label{diffusion:coeff}
\end{equation}
(for large $\nu$).

Using the formula $a=\sqrt{D\tau_{D}}$ let us define diffusion characteristic time
\begin{equation}
\tau_{D}=\frac{a^{2}}{D}\simeq \frac{2a^{2}}{v_{0}^{2}}\nu
\label{diffusion:time}
\end{equation}

More suitable form of the diffusion equation~(\ref{diffusion:equation}) is
\begin{equation}
\Delta N(\vec{r})-\beta^{2}N(\vec{r})=
-\frac{\beta^{2}}{\alpha_{0}}\,\lambda(\vec{r})\,,\quad
\beta^{2}=\frac{2\alpha_{0}\alpha^{2}}{\nu v_{0}^{2}}\,.
\label{diffusion:equation2}
\end{equation}
This equation can be solved using Green functions techniques.

In the present system there are four characteristic times:
\begin{equation*}
\tau_{a}=\frac{a}{v_{0}}\,,\quad
\tau_{\gamma}=\frac{1}{\gamma}\,,\quad
\tau_{\nu}=\frac{1}{\nu}\,,\quad
\tau_{D}\simeq\frac{\tau_{a}^{2}}{\tau_{\nu}}=\nu\tau_{a}^{2}\,,\quad
\end{equation*}
According to initial assumption $\tau_{a}>\tau_{\nu}$ 
(great number of collisions during the flight time)
Also we assume
\begin{equation*}
\tau_{a}<\tau_{\gamma}\,,\quad
\tau_{a}<\tau_{D}\,.
\end{equation*}

\section{Exact solution for infinite region}
\label{sect3}
For spatially infinite region
the equation~(\ref{basic:equation}) can be solved using Fourier transformation with respect to $\vec{r}$:
\begin{equation*}
\hat{\rho}(\vec{k}, \vec{v})
=\bigl[ \hat{\lambda}(\vec{k})+\nu\hat{N}(\vec{k})\bigr]\frac{W(\vec{v})}{\alpha+i\vec{k}\cdot\vec{v}}\,,\quad
\mbox{where}\quad
\hat{\rho}(\vec{k}, \vec{v})
=\int\rho(\vec{r}, \vec{v})\,e^{i\vec{k}\cdot\vec{r}}\,d\vec{r}\,.
\end{equation*}
Since $\hat{N}(\vec{k})=\ds\int\hat{\rho}(\vec{k}, \vec{v})\,d\vec{v}$, 
we get after integration over $\vec{v}$
\begin{equation}
\hat{N}(\vec{k})
=\bigl[\hat{\lambda}(\vec{k})+\nu\hat{N}(\vec{k})\bigr]\hat{F}(\vec{k})\,,\quad
\mbox{where}\quad
\hat{F}(\vec{k})=\int\frac{W(\vec{v})\,d\vec{v}}{\alpha+i\vec{k}\cdot\vec{v}}\,,
\label{basic:fourier}
\end{equation}
so that we can express the function $\hat{N}(\vec{k})$ as
\begin{equation}
\hat{N}(\vec{k})=\frac{\hat{\lambda}(\vec{k})\hat{F}(\vec{k})}{1-\nu\hat{F}(\vec{k})}\,.
\label{basic:N}
\end{equation}
The function $\hat{F}(\vec{k})$ is the Voigt profile and can be presented as follows
\begin{equation}
\hat{F}(\vec{k})
=W_{0}\int\frac{e^{-v^{2}/v_{0}^{2}}\,d\vec{v}}{\alpha+i\vec{k}\cdot\vec{v}}
=\int\limits_{0}^{\infty}\exp\left( -\frac{k^{2}\xi^{2}v_{0}^{2}}{4}-\alpha\xi\right)\,d\xi
\label{basic:Fk}
\end{equation}
This expression is valid for all spacial dimensions. Note that~(\ref{basic:Fk})
can be written in terms of the error function, namely 
$\hat{F}(\vec{k})=\frac{\sqrt{\pi}}{\alpha}\,A e^{A^{2}}\erfc (A)$, where
$A=\frac{\alpha}{|k|v_{0}}$.
For sufficiently small $k$ there exists the following asymptotic expansion (see~\cite{abramovits}):
\begin{equation}
\hat{F}(\vec{k})\simeq
\frac{1}{\alpha}
\left[1-\frac{1}{2}\left( \frac{kv_{0}}{\alpha}\right)^{2}+\frac{3}{4}\left( \frac{kv_{0}}{\alpha}\right)^{4}+\dots\right]\,,\quad
\mbox{where}\quad
\Bigl| \frac{kv_{0}}{\alpha}\Bigr|\simeq \Bigl| \frac{kv_{0}}{\nu}\Bigr|\ll 1\,.
\label{basic:fk-expand}
\end{equation}

Let us find the signal $\S_{\infty}$.
Taking into account the properties of the Fourier transformation we can write:
\begin{equation}
\S_{\infty}(\Delta\omega)
=\int d\vec{x}\,N(\vec{r})\lambda(\vec{r})
=\frac{1}{(2\pi)^{n}}\int d\vec{k}\,\hat{N}(\vec{k})\hat{\lambda}(\vec{k})\,,
\label{basic:infinite-S}
\end{equation}
where $n$ denotes dimension of the space (1, 2 or 3). 
Substituting~(\ref{basic:N}) we get
\begin{equation*}
\S_{\infty}(\Delta\omega)
=\frac{1}{(2\pi)^{n}}\int 
\frac{\hat{\lambda}^{2}(\vec{k})\hat{F}(\vec{k})}{1-\nu\hat{F}(\vec{k})}\,d\vec{k}
\end{equation*}
Due to the factor $\hat{\lambda}^{2}(k)$ 
the value of $S(\Delta\omega)$ is determined by small $k$, so that we can replace $\hat{F}(k)$
by its expansion~(\ref{basic:fk-expand}):
\begin{equation}
\S_{\infty}(\Delta\omega)
=\frac{1}{(2\pi)^{n}\alpha_{0}}\int 
\frac{\hat{\lambda}^{2}(\vec{k})\,d\vec{k}}{1+\frac{k^{2}v_{0}^{2}}{2\alpha\alpha_{0}}}
\label{basic:S-infty}
\end{equation}
The final result depends on space dimension, we will consider partial cases $n=1$ and $n=2$ (the case $n=3$ is unreachable on experiment if $\lambda(\vec{r})=\lambda_{0}e^{-\vec{r}^{2}/a^{2}}$
and is singular if $\lambda(\vec{r})=\lambda_{0}e^{-(x^{2}+y^{2})/a^{2}}$).

\subsection{One dimension}
In this case 
\begin{equation*}
W(v)=\frac{1}{\sqrt{\pi}v_{0}}\,e^{-v^{2}/v_{0}^{2}}\,,\quad 
\lambda(x)=\lambda_{0}e^{-x^{2}/a^{2}}\,.
\end{equation*}
Since $\hat{\lambda}(k)=\sqrt{\pi}a\lambda_{0}e^{-k^{2}a^{2}/4}$, we obtain from~(\ref{basic:S-infty}):
\begin{equation}
\S_{\infty}(\Delta\omega)
\simeq\frac{\lambda_{0}^{2}a^{2}}{2\alpha_{0}}\int\limits_{-\infty}^{+\infty}\frac{e^{-k^{2}a^{2}/2}\,dk}{1+\frac{v_{0}^{2}}{2\alpha\alpha_{0}}\,k^{2}}
=\frac{\pi a\lambda_{0}^{2}}{\sqrt{2}}\,
A\,e^{A^{2}}\erfc(A)\,,\quad 
A^{2}=\frac{\alpha\alpha_{0}a^{2}}{v_{0}^{2}}
=\frac{\alpha\alpha_{0}a^{2}}{2\langle v^{2}\rangle}\,.
\label{partial:1d-infty}
\end{equation}
Since $\langle v^{2}\rangle=v_{0}^{2}/2$.
Using the notations introduced above and~(\ref{partial:1d-infty}) 
we can write $\S_{\infty}$ in the following form:
\begin{equation*}
\S_{\infty}
=\sqrt{\frac{\pi}{2}}\,\frac{a\lambda_{0}^{2}A^{2}}{\alpha_{0}}\,\int\limits_{0}^{\infty}\frac{e^{-A^{2}\xi}\,d\xi}{\sqrt{1+\xi}}\,,
\end{equation*}
or, taking into account the fact that for $\nu\gg\gamma, \Delta\omega$ the combination
$A^{2}=\alpha\alpha_{0}\tau_{a}^{2}\simeq \alpha_{0}\tau_{D}$ and
\begin{equation}
\S_{\infty}=\sqrt{\frac{\pi}{2}}\,a\lambda_{0}^{2}
\int\limits_{0}^{\infty}\frac{e^{-\alpha_{0}\tau}\,d\tau}{\bigl[1+\tau/\tau_{D}\bigr]^{1/2}}\,.
\label{diffusion:profile-1}
\end{equation}
The right hand side can be interpreted as the superposition of profiles with the weight factor $e^{-\alpha_{0}\tau}$.

If $\gamma\tau_{D}\gg 1$ we can expand $\S_{\infty}$
\begin{equation*}
\S_{\infty}=\sqrt{\frac{\pi}{2}}\,a\lambda_{0}^{2}
\int\limits_{0}^{\infty}e^{-\alpha_{0}\tau}\left[ 
1-\frac{1}{2}\frac{\tau}{\tau_{D}}+\dots
\right]\,d\tau
=\sqrt{\frac{\pi}{2}}\,a\lambda_{0}^{2}
\frac{1}{\alpha_{0}}\left( 
1-\frac{1}{\alpha_{0}}\frac{1}{2\tau_{D}}+\dots
\right)
\simeq\sqrt{\frac{\pi}{2}}\,\lambda_{0}^{2}\frac{1}{\alpha_{0}+\frac{1}{2\tau_{D}}}\,.
\end{equation*}
This result is correct for small $\Delta\omega$ (i.~e. $|\Delta\omega|\ll\gamma$) 
and $\gamma\tau_{D}\gg 1$, $\gamma\ll \nu$.

If $\gamma\tau_{D}\sim 1$ we can expand $\S_{\infty}$ directly in powers of $\Delta\omega$.
For small $\Delta\omega$
\begin{equation*}
A^{2}=\alpha_0\alpha\tau_{a}^2\simeq\tau_{D}\bigl( \gamma+i\Delta\omega\bigr)-\tau_{a}^{2}\Delta\omega^{2}\,,
\end{equation*}
so that 
\begin{equation*}
\begin{split}
\frac{1}{\alpha_{0}}\,e^{A^{2}}\erfc(A)=&
\sqrt{\frac{\tau_{D}}{\gamma}}\,e^{\tau_{D}\gamma}\erfc(\sqrt{\tau_{D}\gamma})
+\frac{\tau_{D}}{\gamma}
\left[
-\frac{1}{\sqrt{\pi}}+\sqrt{\tau_{D}\gamma}\left( 1-\frac{1}{2\tau_{D}\gamma}
\right)e^{\tau_{D}\gamma}\erfc(\sqrt{\tau_{D}\gamma})
\right]i\Delta\omega\\
+&\left[
\frac{1}{\sqrt{\pi}}
\frac{\tau_{D}}{\gamma}\left( \frac{\tau_{D}}{2}-\frac{3}{4\gamma}\right)
+\frac{1}{2}\sqrt{\frac{\tau_{D}}{\gamma}}\left( 
\frac{\tau_{D}}{\gamma}-2\tau_{a}^{2}-\tau_{D}^{2}-\frac{3}{4\gamma^{2}}
\right)
\right]\Delta\omega^{2}+\dots\,.
\end{split}
\end{equation*}

\subsection{Two dimensions}
The Maxwell distribution function has the form 
\begin{equation*}
W(\vec{v})=\frac{1}{\pi v^{2}_{0}}\,e^{-v^{2}/v_{0}^{2}}\,,\quad 
\lambda(x)=\lambda_{0}e^{-\vec{r}^{2}/a^{2}}\,,
\end{equation*}
the Fourier transformation if $\lambda(\vec{r})$ 
is $\hat{\lambda}(k)=\pi a^{2}\lambda_{0}e^{-k^{2}a^{2}/4}$, so that
\begin{equation}
\S_{\infty}(\Delta\omega)
=\frac{\pi\lambda_{0}^{2} a^{4}}{2\alpha_{0}}
\int\limits_{0}^{+\infty}\frac{ke^{-k^{2}a^{2}/2}\,dk}{1+\frac{v_{0}^{2}}{2\alpha\alpha_{0}}\,k^{2}}
=\frac{\pi\lambda_{0}^{2}a^{2}}{2\alpha_{0}}\,A^{2}e^{A^{2}}\ei_{1}(A^{2})\,,\quad
A^{2}=\frac{\alpha\alpha_{0}a^{2}}{v_{0}^{2}}=\frac{\alpha\alpha_{0}a^{2}}{\langle v^{2}\rangle}\,,
\label{partial:2d-infty}
\end{equation}
here $\ei_{1}(\dots)$ if the integral exponent of the first order\footnote{definition: $\ds\ei_{1}(x)=\int\limits_{x}^{\infty}\frac{e^{-t}}{t}\,dt$\,.}.

The function $\S_{\infty}$ can be rewritten as
\begin{equation}
\S_{\infty}(\Delta\omega)
=\frac{\pi\lambda_{0}^{2}a^{2}}{2}
\int\limits_{0}^{\infty}\frac{e^{-\alpha_{0}\tau}\,d\tau}{1+\frac{\tau}{\tau_{D}}}\,,\quad
\langle v^{2}\rangle=v_{0}^{2}\,.
\label{diffusion:profile-2}
\end{equation}
with the same meaning as~(\ref{diffusion:profile-1}).

In the case $\gamma\tau_{D}\gg 1$
\begin{equation*}
\S_{\infty}(\Delta\omega)\simeq
\frac{\pi\lambda_{0}^{2}a^{2}}{2}
\frac{1}{\alpha_{0}+\frac{1}{\nu \tau_{a}^{2}}}\,.
\end{equation*}
The combination  $\nu \tau_{a}^{2}$ is the diffusion time for 1-dimensional problem.
Note that in 2-dimensional configuration the real diffusion time defined as $\tau_{D}=a^{2}/D$
and expressed in terms of $\langle v^{2}\rangle$ is different (and can be obtained with replacing $D\to 2D$).

If $\gamma\tau_{D}\simeq 1$ the expansion $\S_{\infty}(\Delta\omega)$ is
\begin{equation*}
\begin{split}
\frac{1}{\alpha_{0}}\,A^{2}e^{A^{2}}\ei(A^{2})&
\simeq
\tau_{D}e^{\tau_{D}\gamma}\ei(\tau_{D}\gamma)+
\frac{\tau_{D}}{\gamma}\,e^{\tau_{D}\gamma}\bigl[
e^{\tau_{D}\gamma}+\tau_{D}\gamma\ei(\tau_{D}\gamma)
\bigr]i\Delta\omega\\
-&\left[ 
\frac{\tau_{D}}{\gamma}e^{2\tau_{D}\gamma}\left( \frac{3\tau_{D}}{2}-\frac{1}{2\gamma}
\right)
+\left( \tau_{a}^{2}-\frac{\tau_{D}^{2}}{2}\right)
e^{\tau_{D}\gamma}\ei(\tau_{D}\gamma)
\right]\Delta\omega^{2}\,.
\end{split}
\end{equation*}

\subsection{Interpretation of Diffusion Equation}
Using inverse Fourier transformation in~(\ref{basic:fourier}) 
and the properties of the convolution we can write:
\begin{equation}
N(\vec{r})=\int\bigl[ \lambda(\vec{r}\pr)+\nu N(\vec{r}\pr)\bigr]\,F(\vec{r}-\vec{r}\pr)\,d\vec{r}\pr
=\int\bigl[ \lambda(\vec{r}-\vec{r}\pr)+\nu N(\vec{r}-\vec{r}\pr)\bigr]\,F(\vec{r}\pr)\,d\vec{r}\pr\,,
\label{diffusion:convolution}
\end{equation}
where
\begin{equation}
F(\vec{r})
=\frac{1}{(2\pi)^{n}}\int\hat{F}(\vec{k})e^{-i\vec{k}\cdot\vec{r}}\,d\vec{k}
=W_{0}\int\limits_{0}^{+\infty}\exp\left( -\alpha\xi-\frac{r^{2}}{\xi^{2}v_{0}^{2}}\right)\,\frac{d\xi}{\xi^{n}}\,,\quad
W_{0}=\frac{1}{(\sqrt{\pi}v_{0})^{n}}\,.
\label{basic:fx}
\end{equation}
is transformation of $\hat{F}(k)$ (with the logarithmic singularity).
Both of the convolutions in~(\ref{diffusion:convolution}) can be approximated by an asymptotic expansion
taking into account the peak if $F(\vec{r})$ at $\vec{r}=0$), because its values are determined 
for~$\vec{r}\pr$ close to~$\vec{r}$. Since
\begin{equation*}
N(\vec{r}-\vec{r}\pr)=
N(\vec{r})+x'_{i}\nabla_{i}N(\vec{r})+\frac{1}{2}\,x'_{i}x'_{j}\nabla_{i}\nabla_{j}N(\vec{r})+\dots
\end{equation*}
we get
\begin{equation*}
\int N(\vec{r}-\vec{r}\pr)\,F(\vec{r}\pr)\,d\vec{r}\pr
=N(\vec{r})\int F(\vec{r}\pr)\,d\vec{r}\pr
+\nabla_{i}N(\vec{r})\int x'_{i}F(\vec{r}\pr)\,d\vec{r}\pr
+\frac{1}{2}\,\nabla_{i}\nabla_{j}N(\vec{r})\int x'_{i}x'_{j}F(\vec{r}\pr)\,d\vec{r}\pr
+\dots
\end{equation*}
Using the fact that $F(\vec{r})$ depends only on $r=|\vec{r}|$ we can conclude that
\begin{equation*}
\int N(\vec{r}-\vec{r}\pr)\,F(\vec{r}\pr)\,d\vec{r}\pr
=N(\vec{r})\int F(\vec{r}\pr)\,d\vec{r}\pr
+\frac{1}{2}\,\Delta N(\vec{r})\int F(\vec{r}\pr)r'{}^{2}\,d\vec{r}\pr
+\dots
\end{equation*}
Using the explicit form~(\ref{basic:fx}) one can obtain after simple calculations:
\begin{equation*}
\int F(\vec{r}\pr)\,d\vec{r}\pr=\frac{1}{\alpha}\,,\quad
\int F(\vec{r}\pr)r'{}^{2}\,d\vec{r}\pr=\frac{2\langle v^{2}\rangle}{3\alpha^{3}}\,,\quad
\langle v^{2}\rangle=\int v^{2}W(\vec{v})\,d\vec{v}\,.
\end{equation*}
Hence, 
\begin{equation}
\int N(\vec{r}-\vec{r}\pr)\,F(\vec{r}\pr)\,d\vec{r}\pr
=\frac{1}{\alpha}
\left[ 
N(\vec{r})+\frac{\langle v^{2}\rangle}{3\alpha^{2}}\,\Delta N(\vec{r})+\dots
\right]
\label{basic:N-convolution}
\end{equation}
The parameter $\bigl| \frac{v_{0}}{\alpha}\bigr|\simeq \frac{v_{0}}{\nu}$ is assumed to be sufficiently small with respect to~$a$.
The same expansion is valid for convolution $F(x)$ with $\lambda(x)$.
Note that the convolution~(\ref{diffusion:convolution}) is the same as~(\ref{basic:operator:form}).

To prove that our approach is correct one can solve the diffusion equation 
and compare results with the previous calculations based on Fourier transformation.

\subsection{One dimension}
The Green function with the boundary conditions $G(x, x')|_{x=\pm \infty}=0$ is 
\begin{equation*}
G(x, x')=\frac{1}{2\beta}\,e^{-\beta|x-x'|}\,,
\end{equation*} 
so that 
\begin{multline*}
N(x)=\frac{1}{2\beta}\int\limits_{-\infty}^{+\infty}e^{-\beta|x-x'|}f(x')\,dx'
=\frac{\beta\lambda_{0}}{2\alpha_{0}}\int\limits_{-\infty}^{+\infty}e^{-\beta|x-x'|-x'{}^{2}/a^{2}}\,dx'\\
=\frac{\sqrt{\pi}\beta a\lambda_{0}}{4\alpha_{0}}
e^{\varepsilon^{2}}\left[ 
e^{-\beta x}\erfc\left( \frac{x}{a}-\varepsilon\right)
+e^{\beta x}\erfc\left( \frac{x}{a}+\varepsilon\right)
\right]\,,
\end{multline*}
where $\varepsilon=\beta a/2$. The integral $\S_{\infty}^{(1)}=\ds\int\limits_{-\infty}^{+\infty}\lambda(x)N(x)\,dx$
can be easily calculated and coincides with~(\ref{partial:1d-infty}).

\subsection{Two dimensions}
The equation~(\ref{diffusion:equation2}) can be solved using the Green function for $R\to\infty$
\begin{equation*}
G(\vec{r})=\frac{1}{2\pi}\,K_{0}(\beta r)\,,
\end{equation*}
Taking into account the boundary conditions we get
\begin{equation*}
N(r)=\int d\vec{r}\,'G(\vec{r}-\vec{r}\,')f(\vec{r}\,')=
\frac{1}{2\pi}\int d\vec{r}\,'K_{0}(\beta |\vec{r}-\vec{r}\,'|)f(\vec{r}\,')\,.
\end{equation*}
The integral over angle variables can be calculated using the expansion
\begin{equation*}
K_{0}\bigl(\beta|\vec{r}-\vec{r}\,'|\bigr)
=I_{0}(\beta r_{<})K_{0}(\beta r_{>})+2\sum\limits_{n=1}^{\infty}I_{n}(\beta r_{<})K_{n}(\beta r_{>})\cos n(\varphi-\varphi')\,,
\end{equation*}
where $r_{>}$ and $r_{<}$
are the greater and smaller value from $r$ and $r'$, 
\begin{equation*}
|\vec{r}-\vec{r}\,'|=\sqrt{r^{2}+r'{}^{2}-2rr'\cos(\varphi-\varphi')}\,.
\end{equation*}
Therefore
\begin{equation}
N(r)=2\pi\int dr'\,r'I_{0}(r_{<})K_{0}(r_{>})\,f(r')\,.
\label{partial:N2-infinite}
\end{equation}
The function $\S^{(2)}_{\infty}$ after simple transformations becomes
\begin{equation}
\S^{(2)}_{\infty}=\frac{4\pi\beta^{2}}{\alpha_{0}}\int\limits_{0}^{\infty}r\,dr\,\lambda(r)
K_{0}(\beta r)\int\limits_{0}^{r}dr'\,r'\lambda(r')I_{0}(\beta r')\,.
\label{partial:profile2-infty}
\end{equation}
It can be shown (using the McDonald integral~\cite{bateman}, 7.7.6 (37)) 
that this integral coincides with~(\ref{partial:2d-infty}).

\section{Diffusion Equation in Finite Region}
\label{sect4}
Other approach to diffusion equation is following: the equation~(\ref{basic:equation})
can be solved formally when the last term in right hand side is treated 
as inhomogeneous term (see~\cite{mors}).  The result is the same as~(\ref{diffusion:equation}).
Namely, in 1-dimensional case, where~(\ref{basic:equation}) is 
\begin{equation*}
v\frac{\partial \rho(x, v)}{\partial x}+\alpha_{0}\rho(x, v)=W(v)\biggl[\lambda(x)+\nu N(x)\biggr]\,,\quad
N(\pm R)=0\,.
\end{equation*}
and $-R\leqslant x\leqslant R$
The total solution reads
\begin{equation*}
\rho(x, v)=C(v)e^{-\alpha x/v}
+\int\limits_{x_{0}}^{x}\frac{W(v)}{v}\,\bigl[\lambda(x')+\nu N(x')\bigr]e^{-\alpha (x-x')/v}\,dx'\,.
\end{equation*}
Therefore the part of solution caused by inhomogeneous term is
\begin{equation*}
N(x)
=\int\limits_{-\infty}^{\infty}dv\,\rho(x, v)\,dv
=\int\limits_{-R}^{+R}dx'\,
\bigl[\lambda(x')+N(x')\bigr]\,F(x-x')\,,
\end{equation*}
$F(x)$ was defined in~(\ref{basic:fx}). 
The homogeneous part of the solution that include 
$C(v)$ vanishes according to initial assumptions due to the presence of the
factors $\rho(\pm R, v)=0$~--- 
the coherence will be lost after collisions with walls $x=\pm R$.

The last integro-differential equation can be written in the form of diffusion equation.
Taking into account the behavior of $F(x)$ and expanding $N(x)$ in series, 
one can obtain with additional assumption
$|\alpha R/v_{0}|\gg 1$ (i.~e. $\nu R/v_{0}=\nu \tau_{R}\gg 1$):
\begin{equation*}
\int\limits_{-R}^{+R}\,
N(x')\,F(x-x')\,dx'=
\frac{1}{\alpha}\left[
N(x)+\frac{\langle v^{2}\rangle}{\alpha^{2}}\,N''(x)+\dots
\right]\,,
\end{equation*}
this coincides with~(\ref{basic:N-convolution}). Therefore, the diffusion equation is valid also for 
finite region when $\nu \tau_{R}\gg 1$.

Let us consider the solution of the diffusion equation 
in the form~(\ref{diffusion:equation2}) for different dimensions.
It appears that most effective for this purpose is the Green functions technique.

\subsection{One dimension}
The equation for $N(x)$ is
\begin{equation}
N''(x)-\beta^{2}N(x)=-f(x)\,,\quad
N(-R)=N(R)=0\,.
\label{diffusion:equation2-1}
\end{equation}
where
\begin{equation*}
f(x)=\frac{\beta^{2}}{\alpha_{0}}\,\lambda(\vec{r})\,,\quad
\beta^{2}=\frac{\alpha_{0}\alpha^{2}}{\nu v_{0}^{2}}=\frac{\alpha_{0}\alpha^{2}}{2\nu\langle v^{2}\rangle}\,.
\end{equation*}
Taking into account the fact that 
both the functions $N(x)$ and $f(x)$ are even we can replace this problem by
\begin{equation}
N''(x)-\beta^{2}N(x)=-f(x)\,,\quad
N'(0)=N(R)=0\,.
\label{diffusion:equation2-1a}
\end{equation}
Let us find the Green $G(x, x')$ function for this equation with the boundary conditions
$\frac{d}{dx}G(x, x')|_{x=0}=0$ and $G(x, x')|_{x=R}=0$. It is easy to find that
\begin{equation}
G(x, x')=\frac{\ch\bigl(\beta x_{<}\bigr)\sh\bigl(\beta (R-x_{>})\bigr)}{\beta\ch\bigl(\beta R\bigr)}\,,
\label{diffusion:green1-finite}
\end{equation}
where $x_{<}$ and $x_{<}$ denote the greater and the smaller values from $x$ and $x'$.

The partial solution of inhomogeneous equation~(\ref{diffusion:equation2-1a}) is
\begin{equation*}
N(x)=\frac{1}{\beta}\frac{\ch\beta x}{\ch\beta R}\int\limits_{0}^{R}\sh\beta(R-x')f(x')\,dx'
-\frac{1}{\beta}\int\limits_{0}^{x}\sh\beta(x-x')f(x')\,dx'\,.
\end{equation*}
Taking into account the boundary conditions we can conclude that this expression is a total solution of~(\ref{diffusion:equation2-1a}).
Let us denote for simplicity
\begin{equation*}
C(x)=\int\limits_{0}^{x}e^{-y^{2}/a^{2}}\ch\beta y\,dy\,,\quad
S(x)=\int\limits_{0}^{x}e^{-y^{2}/a^{2}}\sh\beta y\,dy\,.
\end{equation*}
These integrals can be calculated in terms of error function:
\begin{equation*}
\begin{aligned}
C(x)=&
\frac{\sqrt{\pi}}{4}\,ae^{\varepsilon^{2}}
\left[ 
\erf\left( \frac{x}{a}-\varepsilon\right)
+\erf\left( \frac{x}{a}+\varepsilon\right)
\right]\,, &&
C(\infty)=\frac{\sqrt{\pi}}{2}\,a e^{\varepsilon^{2}}\,,\quad
\\
S(x)=&
\frac{\sqrt{\pi}}{4}\,ae^{\varepsilon^{2}}
\left[ 
\erf\left( \frac{x}{a}-\varepsilon\right)
-\erf\left( \frac{x}{a}+\varepsilon\right)
+2\erf(\varepsilon)
\right]
&&
S^{(1)}(\infty)=\frac{\sqrt{\pi}}{2}\,a e^{\varepsilon^{2}}\erf(\varepsilon)\,.
\end{aligned}
\end{equation*}
where $\varepsilon=\frac{\beta a}{2}\simeq A$.
Taking into account 
$C'(x)=e^{-x^{2}/a^{2}}\ch\beta x$ and $S'(x)=e^{-x^{2}/a^{2}}\sh\beta x$,
the complex function $\S^{(1)}_{R}$ can be rewritten as
\begin{equation*}
\S_{R}^{(1)}=\frac{\beta}{\alpha_{0}}\,\lambda_{0}^{2}
\int\limits_{0}^{R}
\frac{\sh\beta(R-x)}{\ch\beta R}\,e^{-x^{2}/a^{2}}C(x)\,dx\,.
\end{equation*}
It is easy to show when $R\to\infty$
$\S^{(1)}_{R}\to\S^{(1)}_{\infty}$ according to~(\ref{partial:1d-infty}).
Hence, current interpretation of the initial integral equation in terms of 
diffusion~(\ref{diffusion:equation2-1a}) equation is correct
in our approximation.

For $R\gg a$ one can compare $\S^{(1)}_{R}$ with $\S^{(1)}_{\infty}$. 
After simple calculations we can write
\begin{equation*}
\S^{(1)}_{R}=\S^{(1)}_{\infty}-\frac{2\beta\lambda_{0}^{2}}{\alpha_{0}}\,C^{2}(R)\bigl[1-\th(\beta R)\bigr]
\end{equation*}
For large $x$
\begin{equation*}
C(x)\simeq C(\infty)-\frac{\sqrt{\pi}}{2}\,ae^{\varepsilon^{2}}\erfc\left( \frac{x}{a}-\varepsilon\right)
\end{equation*}
therefore
\begin{equation*}
\S^{(1)}_{R}\simeq\S^{(1)}_{\infty}
+\frac{2\pi a^{2}\beta}{\alpha_{0}}\,e^{2\varepsilon^{2}}\,e^{-2\beta R}\,.
\end{equation*}

\subsection{Two dimensions}
The diffusion equation in the region $r<R$
\begin{equation*}
\Delta N(\vec{r})-\beta^{2}N(\vec{r})=-f(\vec{r})\,,\quad N(\vec{r})\bigr|_{r=R}=0
\end{equation*}
can be solved using the Green function
\begin{equation*}
\Delta G(\vec{r})-\beta^{2}G(\vec{r})=-\delta(\vec{r}-\vec{r}\,')\,,\quad G(\vec{r})\bigr|_{r=R}=0\,.
\end{equation*}
where
\begin{equation*}
G(\vec{r}, \vec{r}\,')=\frac{1}{2\pi}
\left[
K_{0}\bigl(\beta|\vec{r}-\vec{r}\,'|\bigr)
-\frac{I_{0}(\beta r)}{I_{0}(\beta R)}\,I_{0}(\beta r')K_{0}(\beta R)\right.
-\left.2\sum\limits_{n=1}^{\infty}
\frac{I_{n}(\beta r)}{I_{n}(\beta R)}\,I_{n}(\beta r')K_{n}(\beta R)\cos n(\varphi-\varphi')
\right]\,.
\end{equation*}
All the terms with $n\ne 0$ vanish in the integrals over angles and the term
with $K_{0}\bigl(\beta|\vec{r}-\vec{r}\,'|\bigr)$ 
can be written as in the case with $R\to\infty$.

The expression with $N(r)$ has the form
\begin{multline}
N(r)=\int_{0}^{R}r' f(r')
\frac{I_{0}(\beta r_{<})\bigl[I_{0}(\beta R)K_{0}(\beta r_{>})-I_{0}(\beta r_{>})K_{0}(\beta R)\bigr]}{I_{0}(\beta R)}\,dr'
\\
=K_{0}(\beta r)\int\limits_{0}^{r}\,dr'\,r'I_{0}(\beta r')f(r')
+I_{0}(\beta r)\int\limits_{r}^{R}\,dr'\,r'K_{0}(\beta r')f(r')
-\frac{K_{0}(\beta R)}{I_{0}(\beta R)}\,I_{0}(\beta r)
\int\limits_{0}^{R}dr'\,r'I_{0}(\beta r')f(r')\,.
\label{partial:N2-finite}
\end{multline}
The function $\S^{(2)}_{R}$ is
\begin{equation*}
\S^{(2)}_{R}=2\pi\int\limits_{0}^{R}N(r)\lambda(r)\,r\,dr\,.
\end{equation*}
Using the symmetry of the integrands $\S_{R}$ can be presented in more compact form:
\begin{equation}
\S^{(2)}_{R}=
\frac{2\beta^{2}}{\alpha_{0}}\,\lambda_{0}^{2}
\int\limits_{0}^{R}re^{-r^{2}/a^{2}}
\frac{K_{0}(\beta r)I_{0}(\beta R)-K_{0}(\beta R)I_{0}(\beta r)}{K_{0}(\beta R)}\I(r)\,dr\,,\quad
\label{partial:profile2-finite}
\end{equation}
where
\begin{equation*}
\I(r)=
\int\limits_{0}^{r}
xI_{0}(\beta x)e^{-x^{2}/a^{2}}\,dx\,,
\quad
\I(\infty)=\frac{a^{2}}{2}\,\exp\left( \frac{\beta^{2}a^{2}}{4}\right)\,.
\end{equation*}
For large $r$ (substituting $r$ instead of $x$ in the integrand):
\begin{equation*}
\I(r)=
\I(\infty)-
\int\limits_{r}^{\infty}
xI_{0}(\beta x)e^{-x^{2}/a^{2}}\,dx
\simeq
\I(\infty)\left[ 
1-\frac{1}{\sqrt{2\pi\beta r}}\,e^{-r^{2}/a^{2}}
\right]\,.
\end{equation*}

Like to 1-dimensional case one can compare $\S^{(2)}_{R}$
with $\S^{(2)}_{\infty}$. The result is:
\begin{equation*}
\S^{(2)}_{R}=\S^{(2)}_{\infty}+2\pi\lambda_{0}^{2}\frac{\beta^{2}}{\alpha_{0}}
\int\limits_{R}^{\infty}
\left[\frac{I_{0}(\beta r)}{I_{0}(\beta R)}-\frac{K_{0}(\beta r)}{K_{0}(\beta R)}\right]re^{-r^{2}/a^{2}}\I(\beta r)\,dr
-\pi\lambda_{0}^{2}\frac{\beta^{2}}{\alpha_{0}}\frac{K_{0}(\beta R)}{I_{0}(\beta R)}\I(\infty)^{2}
\end{equation*}
Since $K_{0}(x)\simeq \sqrt{\frac{\pi}{2x}}e^{-x}$, $I_{0}(x)\simeq \frac{e^{x}}{\sqrt{2\pi x}}$ for $x\gg 1$ we can rewrite the last expression as
\begin{equation*}
\begin{split}
\S^{(2)}_{R}&=\S^{(2)}_{\infty}+4\pi\lambda_{0}^{2}\frac{\beta^{2}}{\alpha_{0}}
\int\limits_{R}^{\infty}\sqrt{rR}\sh\bigl[\beta(r-R)\bigr]e^{-r^{2}/a^{2}}\I(\beta r)\,dr
-\pi^{2}\lambda_{0}^{2}\frac{\beta^{2}}{\alpha_{0}}\,e^{-2\beta R}\I(\infty)^{2}\\
&\simeq
\S^{(2)}_{\infty}
-\lambda_{0}^{2}\frac{\beta^{2}}{\alpha_{0}}
\left(
2\pi e^{\beta R-R^{2}/a^{2}}+\pi^{2}\,e^{-2\beta R}
\right)\I(\infty)^{2}\,.
\end{split}
\end{equation*}

\subsection{Tree dimensions}
Consider the case when $N(\vec{r})$ vanishes on the cylinder of radius $r=R$ bounded by the planes
$z=\pm l$.
The diffusion equation remains the same:
\begin{equation*}
\Delta N(\vec{r})-\beta^{2}N(\vec{r})=-f(\vec{r})\,,\quad N(\vec{r})\bigr|_{r=R}=0\,, 
\quad N(\vec{r})\bigr|_{z=\pm l}=0
\end{equation*}
The Green axially symmetric function satisfies the equation
\begin{equation*}
\Delta G(\vec{r}, \vec{r}\pr)-\beta^{2}G(\vec{r}, \vec{r}\pr)=-\frac{\delta(r-r')\delta(z-z')}{2\pi r}\,,
\quad G(\vec{r}, \vec{r}\pr)\bigr|_{r=R}=0\,, 
\quad G(\vec{r}, \vec{r}\pr)\bigr|_{z=\pm l}=0\,.
\end{equation*}

First consider the case $R\to \infty$.
Expanding $G(\vec{r}, \vec{r}\pr)$ in series of $\phi_{n}(z)=\sin\frac{\pi n}{2l}\,(l-z)$ (with $\phi_{n}(\pm l)=0$)
we get
\begin{equation}
G(\vec{r}, \vec{r})=\frac{1}{2\pi l}
\sum\limits_{n=1}^{\infty}
\sin (k_{n}z)\sin (k_{n}z')
I_{0}(\lambda_{n}r_{<})K_{0}(\lambda_{n}r'_{>})\,,\quad
\mbox{where}\quad
k_{n}=\frac{\pi n}{2l}\,,\quad
\beta^{2}_{n}=\beta^{2}+k_{n}^{2}\,.
\label{partial:green3}
\end{equation}
Finally, the axially symmetrical solution of the diffusion equation is
\begin{equation*}
N(\vec{r})=\int G(\vec{r}, \vec{r})f(\vec{r}\pr)\,d\vec{r}\pr
=\frac{2}{\pi}\sum\limits_{n=1}^{\infty}
\frac{\sin k_{2n+1}}{2n+1}
\int\limits_{0}^{\infty}dr'\,r'
I_{0}(\beta_{2n+1}r_{<})K_{0}(\beta_{2n+1}r_{>})f(r')\,.
\end{equation*}
Note that the integral over $r$ is the same that $N(r)$ for 2-dimensional problem with 
$\beta\to \beta_{2n+1}$ (see~(\ref{partial:N2-infinite})).

The complex signal is
\begin{equation*}
\S^{(3)}_{\infty}=2\pi\iint dr\,dz\,rN(r)\lambda(r)
=\frac{4l}{\pi^{2}} \sum\limits_{n=0}^{\infty}\frac{1}{(2n+1)^{2}}
\int\limits_{0}^{\infty} dr\,r\lambda(r)N(r)\,,
\end{equation*}
or, denoting by $\S^{(2)}_{\infty}(\beta)$ the profile~(\ref{partial:profile2-infty})
for corresponding 2-dimensional case we get
\begin{equation*}
\S^{(3)}_{\infty}[\beta_{n}]=\frac{2l}{\pi^{2}}\sum\limits_{n=0}^{\infty}\frac{\S^{(2)}_{\infty}[\beta_{n}]}{(2n+1)^{2}}\,.
\end{equation*}

\begin{figure}
\begin{center}
\includegraphics[scale=0.8]{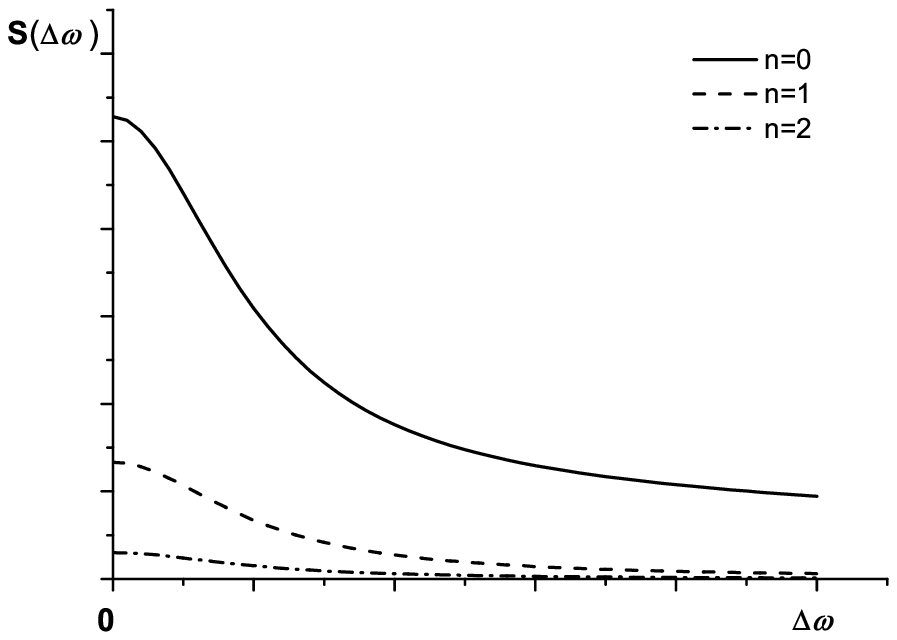}
\includegraphics[scale=0.8]{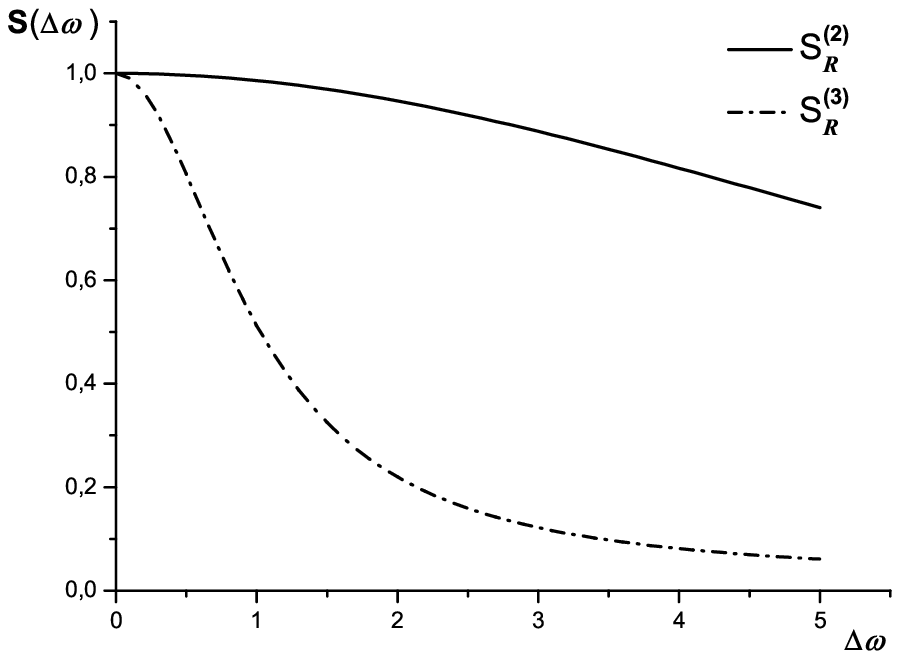}
\caption{Terms in power series $\S^{(3)}_{R}$, comparison of $\S^{(2)}_{R}$ with $\S^{(3)}_{R}$}
\label{fig:3:terms}
\end{center}
\end{figure}

The same procedure can be used for $R<\infty$, the Green function is
\begin{equation}
G(\vec{r}, \vec{r})=\frac{1}{2\pi l}
\sum\limits_{n=1}^{\infty}
\sin (k_{n}z)\sin (k_{n}z')
\bigl[ 
I_{0}(\beta_{n}R)K_{0}(\beta_{n}r_{>})-
I_{0}(\beta_{n}r_{>})K_{0}(\beta_{n}R)
\bigr]
\frac{I_{0}(\beta_{n}r_{<})}{I_{0}(\beta_{n}R)}\,,
\label{partial:green3-finite}
\end{equation}
hence, 
\begin{equation*}
\S^{(3)}_{R}=2\pi\iint dr\,dz\,rN(r)\lambda(r)
=\frac{4l}{\pi^{2}} 
\sum\limits_{n=0}^{\infty}\frac{1}{(2n+1)^{2}}
\int\limits_{0}^{R} dr\,r\lambda(r)N_{R}(r)\,,
\end{equation*}
again, denoting by $\S^{(2)}_{R}(\beta)$ the profile~(\ref{partial:profile2-finite})
for corresponding 2-dimensional case we obtain
\begin{equation*}
\S^{(3)}_{R}[\beta_{n}]=\frac{2l}{\pi^{2}} 
\sum\limits_{n=0}^{\infty}\frac{\S^{(2)}_{R}[\beta_{n}]}{(2n+1)^{2}}\,.
\end{equation*}
In this series only the first terms are significant for result (see fig.~\ref{fig:3:terms}).

\begin{center}
\begin{figure}
\includegraphics[scale=0.7]{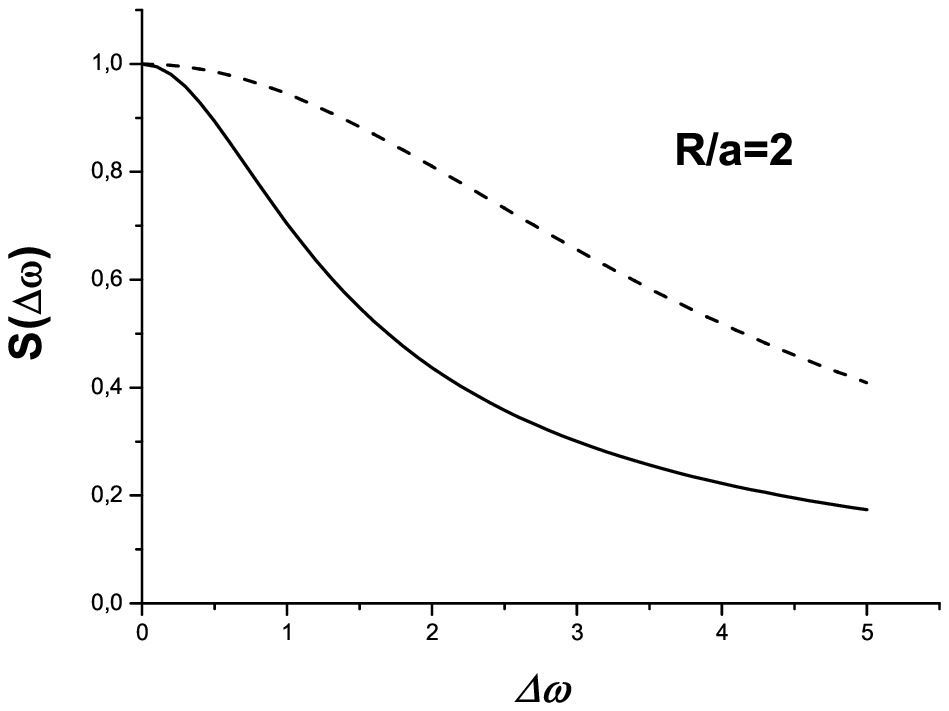}
\includegraphics[scale=0.7]{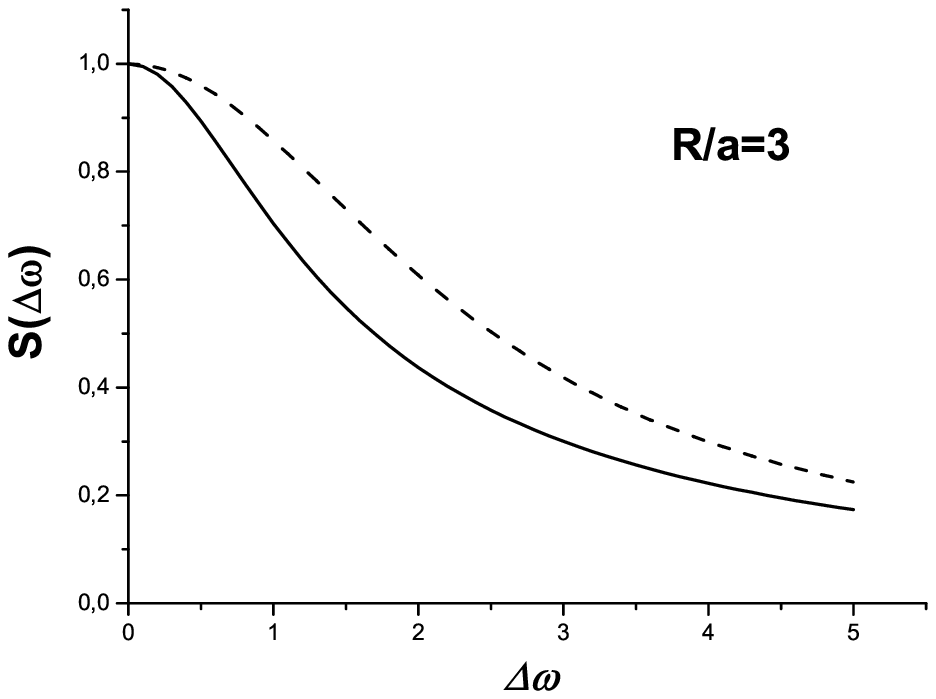}\\
\includegraphics[scale=0.7]{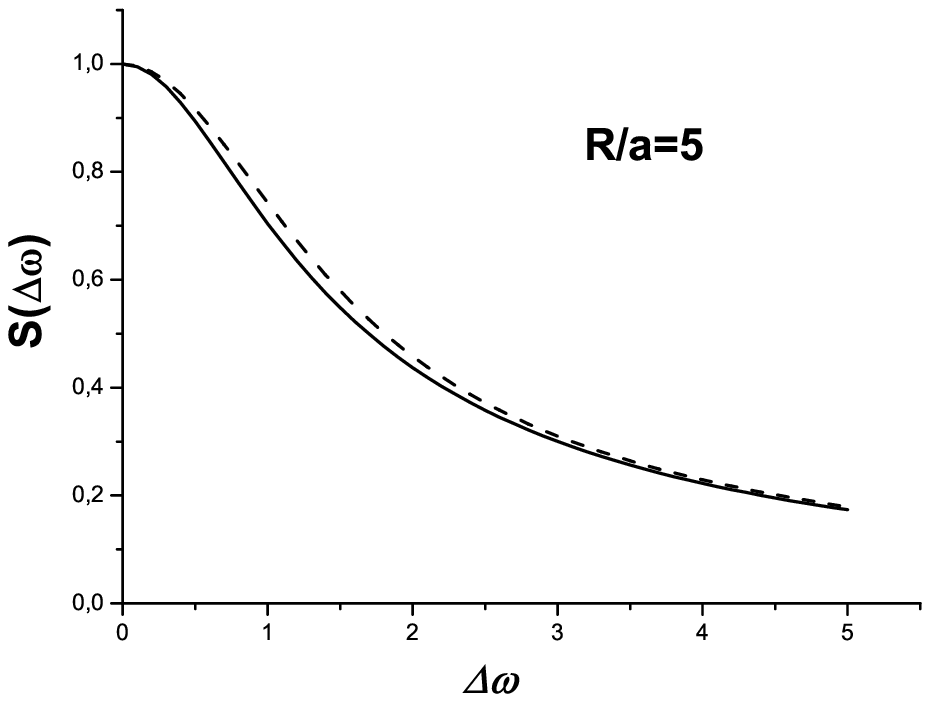}
\includegraphics[scale=0.7]{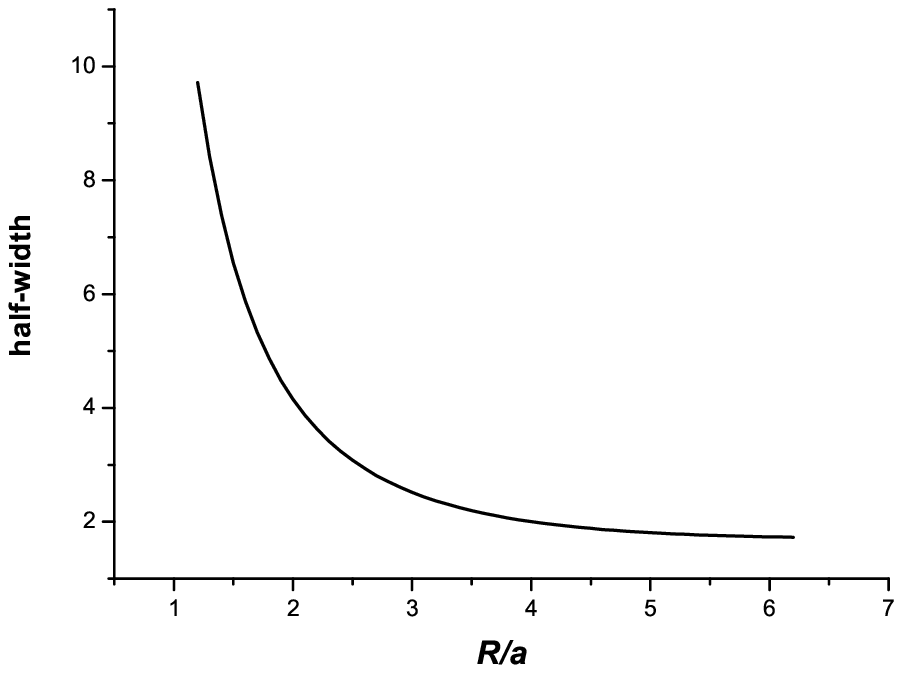}
\caption{The function $S(\Delta\omega)$ for $R=\infty$ and $R<\infty$ (dash), 
half-width as the function of $R$~--- 1-dimensional case}
\end{figure}
\end{center}

\begin{center}
\begin{figure}
\includegraphics[scale=0.7]{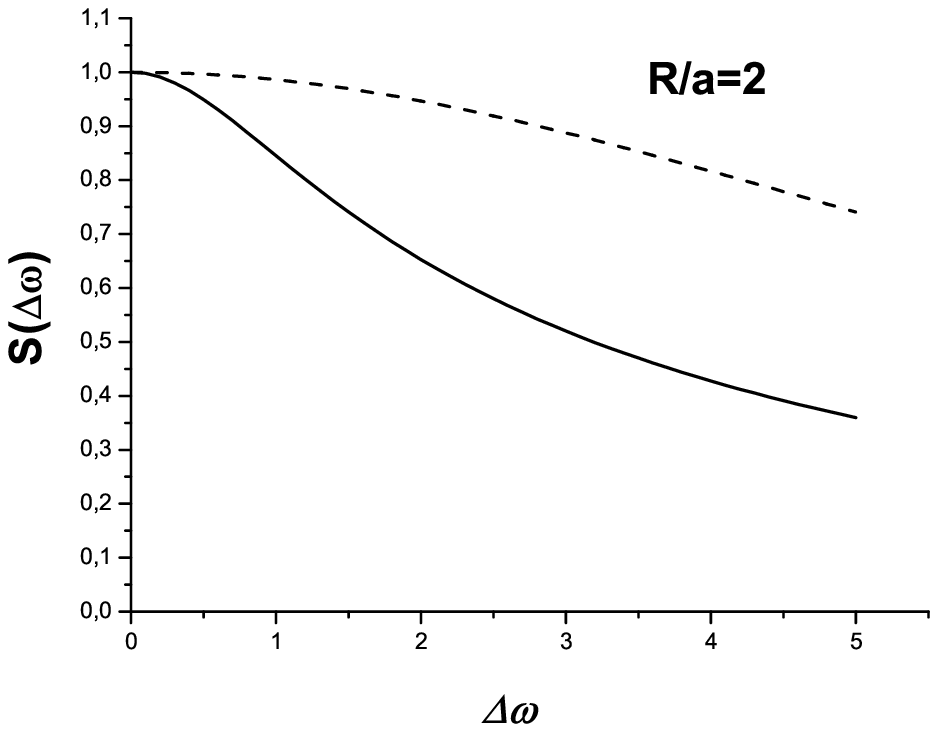}
\includegraphics[scale=0.7]{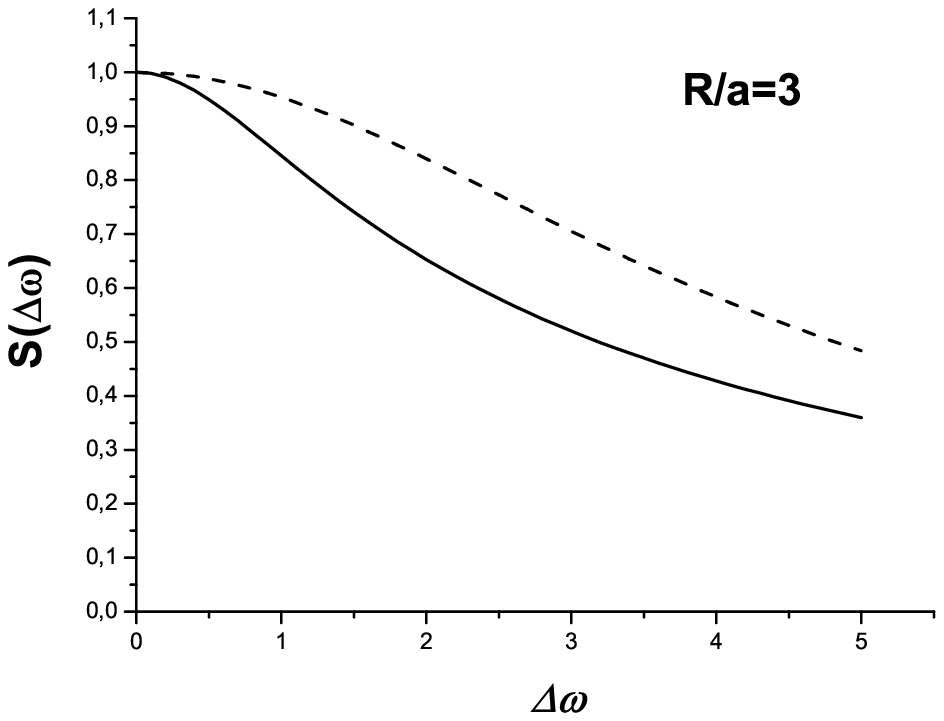}\\
\includegraphics[scale=0.7]{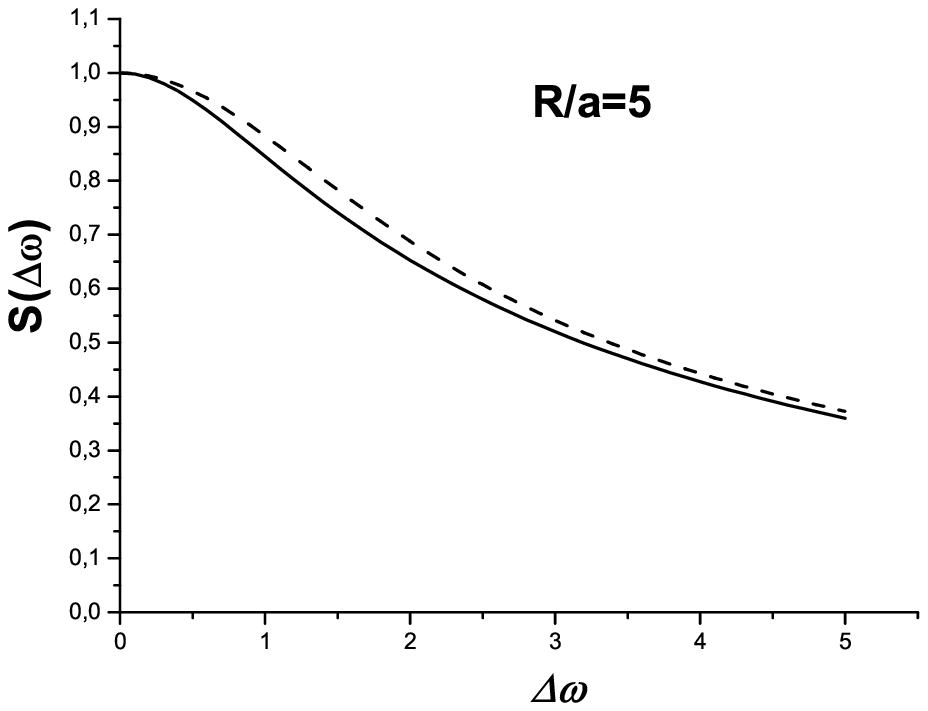}
\includegraphics[scale=0.7]{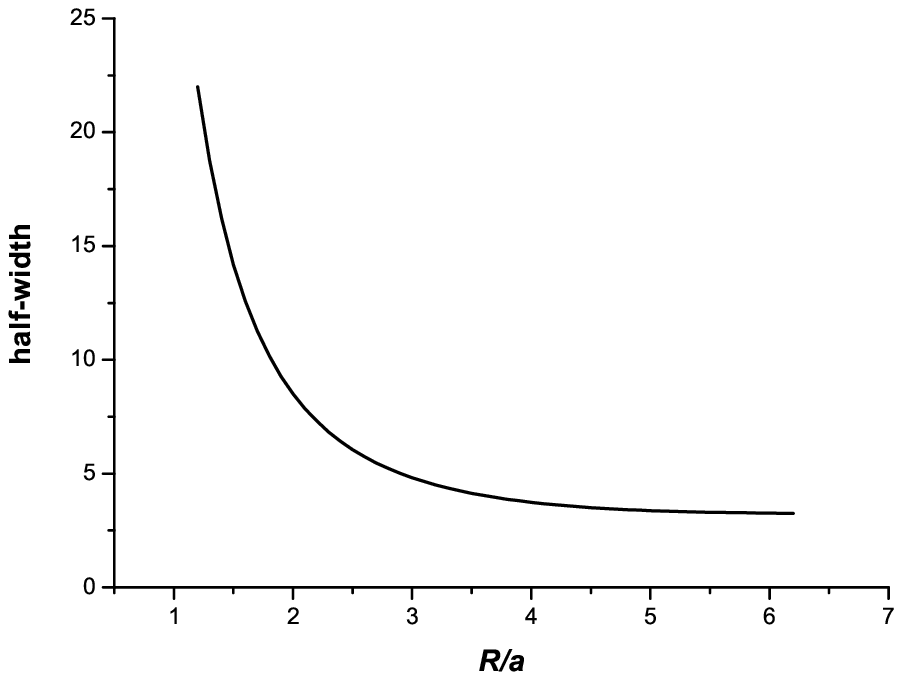}
\caption{The function $S(\Delta\omega)$ for $R=\infty$ and $R<\infty$ (dash), 
half-width as the function of $R$~--- 2-dimensional case}
\end{figure}
\end{center}

\begin{center}
\begin{figure}
\includegraphics[scale=0.7]{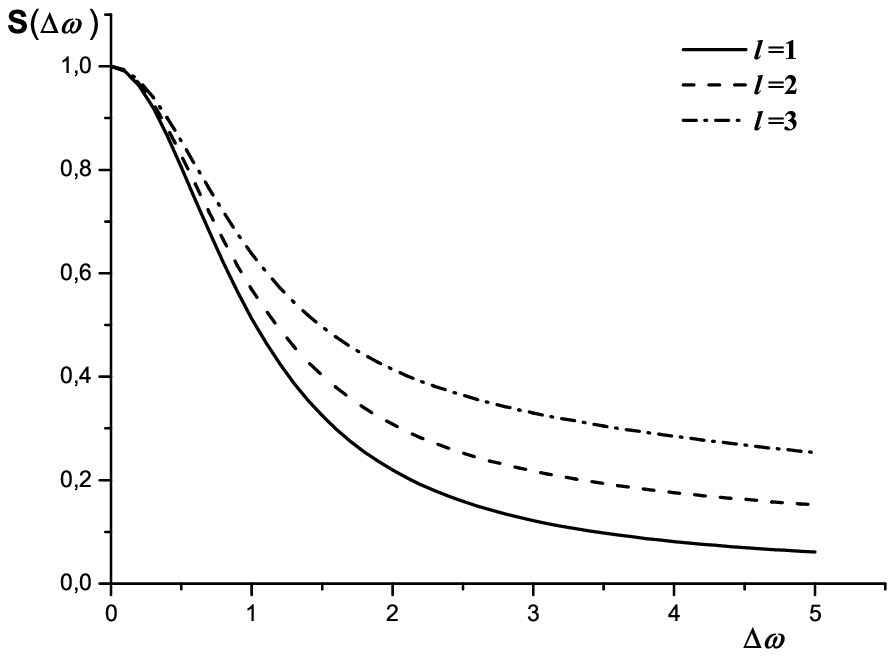}
\includegraphics[scale=0.7]{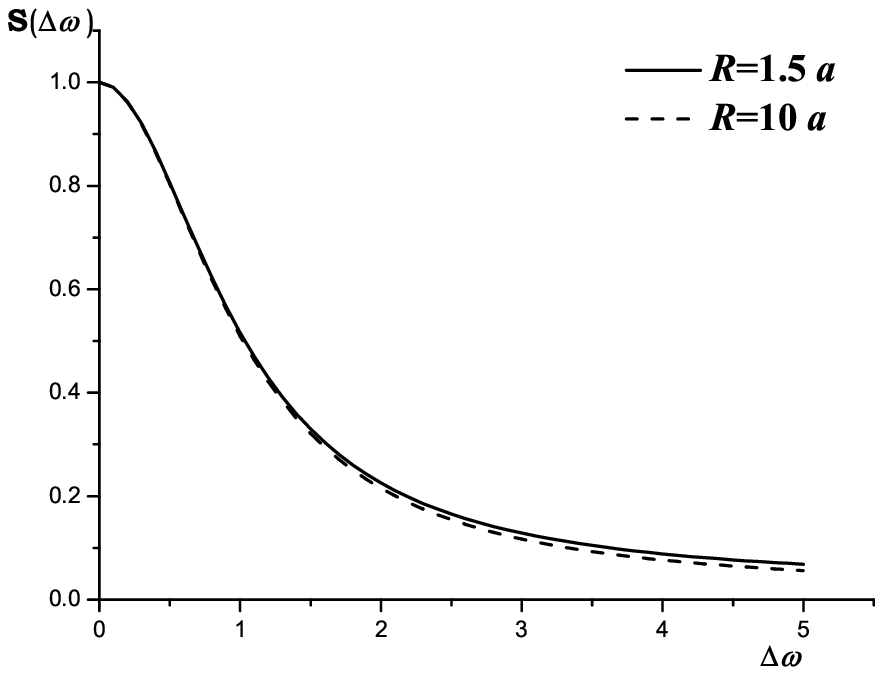}
\caption{The function $S(\Delta\omega)$ for $R<\infty$
with fixed $R$ and $l$~--- 3-dimensional case}
\end{figure}
\end{center}

\section{Discussion}
\label{sect5}
The graphics of $\S(\Delta\omega)$ 
with fixed parameters
\begin{equation*}
\nu=1000\,,\quad
\gamma=1\,,\quad
v_0=100\,,\quad
a=1\,.
\end{equation*}
are presented at the figures.
Characteristic times are
\begin{equation*}
\tau_{a}=0{.}01\,,\quad
\tau_{\gamma}=1\,,\quad
\tau_{D}=0{.}1\,.
\end{equation*}

The values of $R$ (numerically coincide with $R/a$) are
$R=2{.}0\,, 3{.}0\,, 5{.}0$.
All the functions are normalized.
One can see that in the signal in 2-dimensional case is broader
than in 1-dimensional.




\begin{acknowledgments}
We acknowledge support by CNES INTAS and NSAU (06-1000024-9075).
\end{acknowledgments} 


\newpage
\bibliography{paper}

\begin{thebibliography}{7}
\expandafter\ifx\csname natexlab\endcsname\relax\def\natexlab#1{#1}\fi
\expandafter\ifx\csname bibnamefont\endcsname\relax
  \def\bibnamefont#1{#1}\fi
\expandafter\ifx\csname bibfnamefont\endcsname\relax
  \def\bibfnamefont#1{#1}\fi
\expandafter\ifx\csname citenamefont\endcsname\relax
  \def\citenamefont#1{#1}\fi
\expandafter\ifx\csname url\endcsname\relax
  \def\url#1{\texttt{#1}}\fi
\expandafter\ifx\csname urlprefix\endcsname\relax\def\urlprefix{URL }\fi
\providecommand{\bibinfo}[2]{#2}
\providecommand{\eprint}[2][]{\url{#2}}

\bibitem[{\citenamefont{Xiao and et~al}(2006)}]{xiao}
\bibinfo{author}{\bibfnamefont{Y.}~\bibnamefont{Xiao}} \bibnamefont{and}
  \bibinfo{author}{\bibnamefont{et~al}}, \bibinfo{journal}{Phys.\ Rev.}
  \textbf{\bibinfo{volume}{96}}, \bibinfo{pages}{043601}
  (\bibinfo{year}{2006}).

\bibitem[{\citenamefont{Ramsey}(1956)}]{ramsey}
\bibinfo{author}{\bibfnamefont{N.~F.} \bibnamefont{Ramsey}},
  \emph{\bibinfo{title}{Molecular beams}} (\bibinfo{publisher}{Clarendon},
  \bibinfo{address}{Oxford}, \bibinfo{year}{1956}).

\bibitem[{\citenamefont{Rautian and Sobelman}(1966)}]{sobelman}
\bibinfo{author}{\bibfnamefont{S.~G.} \bibnamefont{Rautian}} \bibnamefont{and}
  \bibinfo{author}{\bibfnamefont{I.~I.} \bibnamefont{Sobelman}},
  \bibinfo{journal}{Usp. Phys. Nauk} \textbf{\bibinfo{volume}{90}},
  \bibinfo{pages}{209} (\bibinfo{year}{1966}).

\bibitem[{\citenamefont{Rautian}(1991)}]{rautian}
\bibinfo{author}{\bibfnamefont{S.~G.} \bibnamefont{Rautian}},
  \bibinfo{journal}{Usp. Phys. Nauk} \textbf{\bibinfo{volume}{161}},
  \bibinfo{pages}{151} (\bibinfo{year}{1991}).

\bibitem[{\citenamefont{Abramovitz and Stegun}(1964)}]{abramovits}
\bibinfo{author}{\bibfnamefont{M.}~\bibnamefont{Abramovitz}} \bibnamefont{and}
  \bibinfo{author}{\bibfnamefont{I.}~\bibnamefont{Stegun}},
  \emph{\bibinfo{title}{Handbook of Mathematical Functions}}
  (\bibinfo{publisher}{National bureau of Standards}, \bibinfo{address}{New
  York}, \bibinfo{year}{1964}).

\bibitem[{\citenamefont{Bateman and Erd\'elyi}(1953)}]{bateman}
\bibinfo{author}{\bibfnamefont{H.}~\bibnamefont{Bateman}} \bibnamefont{and}
  \bibinfo{author}{\bibfnamefont{A.}~\bibnamefont{Erd\'elyi}},
  \emph{\bibinfo{title}{Higher Transcendental Functions, vol. 2}}
  (\bibinfo{publisher}{McGray-Hill book company}, \bibinfo{address}{New York},
  \bibinfo{year}{1953}).

\bibitem[{\citenamefont{Mors and Feshbach}(1953)}]{mors}
\bibinfo{author}{\bibfnamefont{F.~M.} \bibnamefont{Mors}} \bibnamefont{and}
  \bibinfo{author}{\bibfnamefont{H.}~\bibnamefont{Feshbach}},
  \emph{\bibinfo{title}{Methods of Theoretical Physics, part 2}}
  (\bibinfo{publisher}{McGray-Hill book company}, \bibinfo{address}{New York},
  \bibinfo{year}{1953}).

\end{thebibliography}

\end{document}